# Hyper-automation-The next peripheral for automation in IT industries.

Ayush Singh Rajput

Student, Amity University, Noida

Dr. Richa Gupta

Asst. Professor, Amity University, Noida

**ABSTRACT**

The extension of legacy business process automation beyond the bounds of specific processes is known as hyperautomation. Hyperautomation provides automation for nearly any repetitive action performed by business users by combining AI tools with RPA. It automates complex IT business processes that a company's top brains might not be able to complete. This is an end-to-end automation of a standard business process deployment. It enables automation to perform task digitalization by combining a brain computer interface (BCI) with AI and RPA automation tools. BCI, in conjunction with automation tools, will advance the detection and generation of automation processes to the next level. It allows enterprises to combine business intelligence systems, address complex requirements, and enhance human expertise and automation experience. Hyperautomation and its importance in today's environment are briefly discussed in this paper. The article then goes on to discuss how BCI and sensors might aid Hyperautomation. The specific sectors of solicitations were examined using a variety of flexible technologies associated to this concept, as well as dedicated workflow techniques, which are also diagrammatically illustrated. Hyperautomation is being utilized to improve the efficiency, accuracy, and human enhancement of automated tasks dramatically. It incorporates a number of automated tools in its discovery, implementation, and automation phases. As a result, it's well-suited to integrating cutting-edge technologies and experimenting with new methods of working.

Keywords- Hyperautomation, Brain computer Interface (BCI), Technology, Used case, Sensors, Industries.

- **INTRODUCTION**

Hyper-automation is the combination of the additional intelligence components of process automation tools and techniques which are essential for growing automation initiatives. This process is helpful in minimizing the manual human effort by implementing automation in the business process. Those who have expertise in Business Process build a clear concept of the requirement of automation rather than other existing employees.

The concept of automation begins with the implementation of Robotic Process Automation (RPA) which extends its peripheral by the involvement of various advanced techniques of Artificial Intelligence along with iBPM and NLP. Implementation of all these technologies evolves end-to-end process designing, automation and monitoring for deployment of impactful initiatives.

As the name suggests, it is useful in lowering the effort of human beings and at the same phase efficiently increasing productivity with improved IT- business automation. One of beneficiary importance of hyper-automation is that it lowers the risk with improved security and will generate an impact if it is used alongside advanced tactics of Artificial Intelligence. Hyper-automation in collaboration with Brain Computer Interface (BCI) will be a booming bull in the field of IT–Industries automation.

Hyper-automation is evaluating with the tremendous development in IT-techniques and technology. Most of the companies are heading towards lowering human efforts by automating their tasks, which will help in the growth of their revenue.

The following research questions are addressed in this article.

RQ1: Research hyper automation and its use.

RQ-2 The function of sensors and flexible technologies in the deployment of hyper-automation will be discussed.

RQ-3 To discuss the concept of a brain-computer interface and a data lake for hyper automation advancement.

RQ-4 To discuss the dedicated hyper automation workflow process.

RQ-5 Difference between Manual, Automated & Hyperautomated.

RQ-6 To discover and discuss hyper-automation's capabilities in the IT industry.

- **Hyper-automation**

The phase hyper-automation was given by *Gartner,* an IT research and advisory group in 2019. *Gartner*, Top list of 10 strategic Technology trends for 2021 isn't the first year that Gartner included hyper automation at the top of its list. Hyper-automation refers to a combination of automation tools with multiple ML applications and packaged software used to free employees from doing repetitive and low-value tasks, allowing them to concentrate on more valuable things. Fig. 1 reflects the variety of versatile technologies associated with the concept of hyper-automation. They are identifying what works to automate, choosing the appropriate automation tools, and extending their capabilities using various technologies of AI and machine learning.

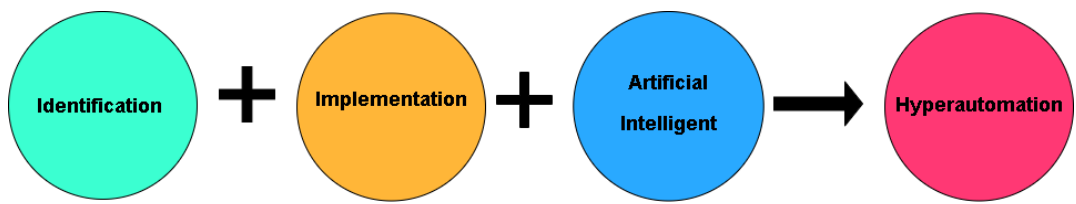

Fig: -1- **A definition to hyperautomation.**

The intention of hyper-automation is to reap the most of the data recorded and created by digitalized operations. It also reduces costs and boosts productivity to gain efficiency through automating automation. The process of hyper automation starts with RPA at its core and expands the automation horizon with AI, process mining, analytics, and other advanced tools. End-to-end innovation and corporate, automation, and monitoring are achievable respect to the implementation of all of these technologies, led to significantly increased value and performance. It overcomes with some e idea of a digital twin of the organization (DTO) which is a virtual representation of how business processes work. Organizations may create better customer experiences while reducing operating costs and increasing profitability by integrating automation and human together.

Although hyper automation is an end-to-end automation process, it's essential to keep in mind that hyper automation is not deliberate to replace humans altogether. Only humans are capable of thinking beyond the frontier.

- **Why Hyper-automation?**

Hyper-automation is required to automate more knowledge labor and to incorporate everyone in a company. It integrates multiple components of process automation, combining tools and technology to improve labor automation. Hyper-automation is the process of combining many machine learning and automation techniques to complete jobs. This is becoming increasingly automatic, expanding the number of people who can benefit from this advanced technology. It refers to a range of instruments as well as the sophistication or phases of automation. It is also critical for hyper automation to select a system that will communicate well with employees. Today's teams are made up of people with a wide range of skills and experiences, so finding a tool that everyone can use and interact with is critical. The selection of a solution narrows the field, potentially giving businesses a significant advantage in automation. Hyper-automation can overcome some of the limitations of a single automation device technique by combining various automation technologies. This enables businesses to go beyond the limitations of each process and automate practically any time-consuming and scalable operation [1,2].

Hyper automations have the potential to unify company strategy, with the end goal of creating and optimizing end-to-end procedures that enable inventive new business concepts. By reducing human interaction in repetitive, time-consuming procedures, the organization improves efficiency, productivity, and moral standards. Organizations must comprehend the scope and role of digital technology in their present workflows. The ever-growing and evolving product market presents another key challenge. It can be difficult to decide which organizations of things should make their consumers accessible. Given the competitive nature of the business, a succession of mergers and acquisitions are expected to decrease redundancies across their goods and improve customers' evaluation of potential suppliers. Hyper-automation has the potential to improve the healthcare industry by providing a better patient experience, more dependable findings, and more accurate data [3,4].

- **Role of Brain Computer to enhance hyper automation**

A brain-computer interface (BCI) is a link between a brain and a platform that allows brain signals to regulate external activity. The interface allows for real - time communication between the brain and the controlled object. In the situation of cursor control, for example, the signal is being sent directly from the brain to the device that directs the cursor, rather than passing through the body's neuromuscular system on its path from the brain to the mouse finger. BCI has the potential to alter all aspects of life by allowing humans to communicate directly with technology.

The majority of BCI technologies are still in their infancy. As a consequence, BCI, as well as signals recognized by EEG sensors, will play an essential part in hyperautomation. It will be "Booming Bull" if the BCI is correctly deployed with automation tools.

- **Role of Sensors to enhance hyper-automation**

Sensors play a big role in automation and Industry 4.0. In a number of ways, such as linear or angular positioning, tilt sensing, leveling, shock, or fall detection, motion, environmental, and vibration sensors are used to monitor equipment health. Microelectromechanical systems or moisture-sensitive EEG sensors (MEMS). Smart sensors have progressed to deliver unprecedented levels of intelligence and communication capabilities, extending the life of older industrial equipment while maximizing the benefits of IoT and cloud computing. These smart sensors have been rapidly upgraded in the market. As a result of greater industrial automation, real-time monitoring of machine parameters, and equipment developed with Predictive Maintenance in mind, sensor demand has expanded globally.

Hyperautomation combines robotic intelligence into the traditional automation process, resulting in an increased work efficiency, speed, and accuracy. The technology can automate practically any repetitive job by integrating BCI, AI with RPA. It automates the automation by recognizing business processes and implementing bots to automate them. It involves the use of a wide range of technologies, which means that firms investing in it must have the necessary tools in place, all of which must be interoperable. Using AI and machine learning, hyperautomation builds "digital twins," or virtual replicas of processes or actual assets. Digital twins are monitored by network-connected sensors and other devices, which collect mountains of data on their state and condition [13,14].

- **Technologies Associated with Hyper-Automation**

| Discovery phase | Implementation of Automation | Automation with AI |
|---|---|---|
| The discovery phase, also known as the scoping phase, which includes process mining, text mining & process analysis. In this phase, a hyper automation tool's design focuses on procedures that require automation and on strategic goals. | This phase also refers to the final stage of the solution's transition from development to manufacturing. This process may be referred to as deployment, go-live, rollout, or installation, depending on your project. | The terms artificial intelligence (AI) and automation are frequently confused and used interchangeably. |

| Discovery Phase involves different areas, they are as: <br><br> 1. Process Mining. <br> 2. Task Mining. <br> 3. Process Analysis. | RPA, PAAS, Workload automation, and Business logic technologies are all included in this phase. | Automation can be used in conjunction with AI such as machine learning and deep learning to achieve even better results in a process we call "AI automation." AI automation is powerful because it combines the advantages of business process automation—increased speed, efficiency, time savings, and scalability—with AI technology's insights, flexibility, and processing capacity. AI automation enables companies to improve their capabilities while delegating mundane tasks to machines. [05] |
|---|---|---|

- **Dedicated workflow process for Implementation of Hyper-automation**

The hyper-automation methodology's workflow is depicted in Figure 6. This begins with a thought that arises in the human brain, which is then followed by information extracted by EEG tools, which is then followed by Brain computer interface (BCI) methodology, which is further followed by layer two of micro processing flow for "Process mining & Task mining," which allows for the analysis of these processes for the extraction of knowledge from the information, and then layer one of micro processing flow for information extraction. After the Micro process flow is completed, it is forwarded to the next level, which is AI-assisted automation. Different forms of machine learning and deep learning tools are bundled together to start automation through various model training algorithms, which are then validated and double-checked. Finally, the machine-readable enhanced data is completed and confirmed. This results in output data, which is either desirable data, if not then in that case the procedure is repeated.

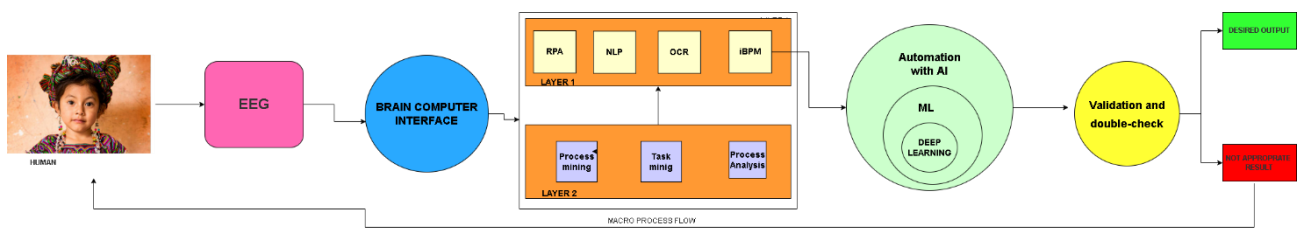

Figure-6- **Dedicated workflow diagram for implementation of hyper-automation.**

Hyperautomation makes it easier to bridge organizational boundaries since it offers more complete automation options. Many areas of company decision-making can be automated with it. Rapid process automation, enhanced analytical applications, greater employee happiness and motivation, value-added workforce labor, more accurate insights, better compliance and risk management, improved productivity, and improved teamwork are just a few examples [06-07]. Companies are focusing on attaining end-to-end process automation as they seek to minimize functional and process constraints by leveraging a wider range of automation solutions. Both the client and the employee benefit from hyperautomation, and both are proportionate to the company in the end. Business and IT stakeholders utilize a process modeling tool to capture, preserve, and optimize processes. It is connected to and prioritized in all areas where there is room for development. Managers and financial teams can use decision-modeling software to automate manual inspections until the entire process is automated. Process automation and human enhancement, as well as commitment, are all driven by AI. A comprehensive toolkit for intelligent automation allows for better, smarter, and faster results by focusing on each process stage. Various new technologies combine with hyperautomated technology to assist organizations in achieving maximum end-to-end business [08-09].

Hyperautomation helps consumers save time and money while reducing errors. It's been utilized to boost customer service and meet management goals. Robotic arms that do jobs more swiftly and with fewer errors are examples of automation. Hyperautomation for intelligent tasks, on the other hand, is the use of a robotic brain. It refers to a wide variety of artificial intelligence (AI) technology. When these technologies are integrated with automation software, the potential for improved flexibility is considerably increased. As previously said, hyperautomation technology is wide and continually expanding. It's a set of business practices, designs, and judgments that apply to a variety of technologies with varied degrees of automation.

- **Difference between Manual, Automated & Hyperautomated.**

| Parameter | Hyperautomated | Automated | Manual |
|---|---|---|---|
| **Technologies Associated** | **Carried out using a variety of machine learning, packaged software, and automation technologies.** | **Carried out by automated tools.** | **Carried out through direct involvement of human being.** |
| **Outcome** | Efficient as well as smart operations. | Efficient outcome. | Accurate and risk free operation. |
| **Process Timing** | Hyperautomated process is significantly faster than a automated approach. | A manual approach is much slower than automated process. | Manual testing requires a lot of time and manpower. |

| **Initial Investment** | Hyperautomated process requires a larger initial cost. Even though over time, ROI is superior. | The automated procedure requires a lesser initial investment then hyperautomated process. In the long run, ROI is lower than Hyperutomation processing. | Manual process requires a smaller initial outlay of funds. In the long term, ROI is lower than automation processing. |
|---|---|---|---|
| **Cost Efficient** | Cost effective for high volume regression, but not for lower volume. | Not cost effective for lower volume. | Cost effective for lower volume. |
| **Degree of Coverage** | All-encompassing: "Everything that can be automated will be automated." | What procedures can we automate, if appropriate? | All-process involves human envolvement. |
| **Scope** | Is a system of platforms, technologies, and related resources. | Is relevant from a single platform. | Is conducted from muntiple platform. |

- **Discover of Hyperautomation's capabilities in the IT industry.**

Hyperautomation is a technique or framework that allows numerous automation technologies to be used continuously and simultaneously. They include task identification, the capacity to reuse automated procedures with agility, and their capabilities. Hyper-automation aims to cut costs, enhance productivity, increase efficiencies, and make better use of the digital process data that is generated and retrieved. Organizations might be able to use this data to make better, timelier business decisions. Hyperautomation is a platform that helps businesses expand, integrate, and improve their internal automation. It examines and expands on the advantages and disadvantages of RPA tools. Hyperautomation is unique among automation frameworks in that it focuses solely on the creation of automation tools or concepts.

Business automation, repeat activity automation, human competency enhancement, speeding up the digital journey, operational scalability, and billing cycle automation are all feasible with automation. Increase operational efficiency, gain flexibility, automate workload, detect risk, and operate more efficiently. All of these factors combine to create an end-to-end automation process that boosts an organization's productivity and efficiency [10]. The efficiency of the company's current IT infrastructure and business operations is critical to automation. Robotic process automation combines machine learning, packaged software, and work automation technologies to merge traditional systems and hyper-automation. To improve the lives of employees, higher outcomes and more productivity are required. In the industrial industry, several digital technologies are used to achieve automation [11-12]. Automation has been employed for many years over the world; nevertheless, with daily changing technological breakthroughs, it has become necessary to adapt to hyperautomation.

- **Analysis of the research**

More than merely putting in place adequate task management systems is required for hyper-automation. It also demands human participation. This is because people are the ones who make the decisions, and they can use technology to evaluate facts and apply logic. In order to develop reports and extract data from social media, a business can rely on the tools to apply machine learning to obtain customer sentiments. It means that low-value occupations should be performed utilizing automation technologies, advanced artificial intelligence, and machine learning to generate outputs and run productively with minimal human intervention. In combination with humans, hyperautomation might provide a constantly trained, flexible, and ready working environment that makes quick and precise judgments based on data and insights. Employees can learn the most up-to-date company and market facts through hyperautomation, allowing them to do their jobs flawlessly.

Hyperautomation is a term that refers to the use of a combination of automation technologies to improve and expand human abilities. It means that low-quality operations are carried out properly using automated tools, machine learning, and sophisticated artificial intelligence to generate outputs and operate efficiently without the need for human engagement. Hyperautomation has the potential to create a workplace that is well-informed, adaptable, and capable of making quick, precise decisions based on data and insights. Model recognition is used to determine what to do next and to optimize operations with the least amount of human input. The system's algorithm is taught using training data and then utilized to create a model. Hyperautomation's ability to create a confluence of various types of automation that seamlessly complement one another at the highest value is a critical component.

- **Conclusion**

To tackle complex challenges and optimize operations, emerging technologies such as brain computer interface and artificial intelligence (AI) are coupled with automation. Hyperautomation has the ability to bring people together by empowering technology and people to work side by side and together. It utilizes technology to analyze humongous data and apply insights to its company as effective policymakers; hyper-automation alters enterprises by optimizing corporate processes by reducing repetitive operations and automating manual ones. Data lakes, on the other hand, help firms to perform activities with consistency, precision, and speed. It keeps all company-related data in both unstructured & structured form, making it easier to analyze the data and retrieve the necessary information quickly. Prices are lower as a result, and customer service is usually better. Any innovative approach to corporate operations or infrastructure can be complicated by hyperautomation. Hyperautomation automates a wide range of instruments, making it possible to tackle complex problems quickly. Several businesses have invested in cutting-edge technology to overcome limitations. Hyperautomation enables a business to automate activities automatically in order to boost productivity and provide more value to customers. Advanced automation that appears to complete activities and procedures faster, more efficiently, and with fewer errors is known as hyperautomation. The anticipated outcomes and core value requirements, such as revenue, cost savings, and risk management, are met as a result.

- **Competing Interests Declaration**

There are no vested interests to contend with.